Multiband ballistic transport and anisotropic commensurability magnetoresistance in antidot lattices of AB-stacked trilayer graphene

Shingo Tajima[1], Ryoya Ebisuoka[1], Kenji Watanabe[2], Takashi Taniguchi[2] and Ryuta Yagi[1]

[1]Graduate School of Advanced Sciences of Matter (AdSM), Hiroshima University, Higashi-Hiroshima 739-8530, Japan,
[2]National Institute for Materials Sciences (NIMS), Tsukuba 305-0044, Japan.

ABSTRACT

Ballistic transport was studied in a multiple-band system consisting of an antidot lattice of AB-stacked trilayer graphene. The low temperature magnetoresistance showed commensurability peaks arising from matching of the antidot lattice period and radius of cyclotron orbits for each mono- and bilayer-like band in AB stacked trilayer graphene. The commensurability peak of the monolayer-like band appeared at a lower magnetic field than that of the bilayer-like band, which reflects the fact that the Fermi surface of the bilayer-like band is larger than that of monolayer-like band. Rotation of the antidot lattice relative to the crystallographic axes of graphene resulted in anisotropic magnetoresistance, which reflects the trigonally warped Fermi surface of the bilayer-like band. Numerical simulations of magnetoresistance that assumed ballistic transport in the mono- and bilayer-like bands approximately reproduced the observed magnetoresistance features. It was found that the monolayer-like band significantly contributes to the conductivity even though its carrier density is an order smaller than that of the bilayer-like band. These results indicate that ballistic transport experiments could be used for studying the anisotropic band structure of multiple-band systems.

(Corresponding author: yagi@hiroshima-u.ac.jp)

# 1. Introduction

Since the discovery of graphene [1] a number of efforts have been made to clarify its electronic properties [2-4]. Electrons in monolayer graphene were found to have the character of massless Dirac Fermions [1]. Theory predicts that the band structure evolves with the number of layers; *i.e.*, graphene has a different band structure depending on the number of layers [5-19]. In particular, AB-stacked graphene is expected to show an even-odd layer number effect; multilayer graphene with an even number of layers, $2N$, has $N$ bilayer-like bands; multilayer graphene with an odd number of layers, $2N + 1$, has $N$ bilayer-like bands and one monolayer-like band. [5,10-16]. To date, the band structures of graphene with various numbers of layers have been measured in experiments. High energy band structure was observed measured via optical spectroscopic measurements in the early stage of graphene research [20-26]. Low energy band structure, which is important for transport measurements, has been measured by the use of Shubnikov-de Haas (S-dH) oscillations in high quality graphene samples [1,16,27-39]

Because graphene's crystallographic structure is a honeycomb lattice of carbons, the band structure is more or less anisotropic (trigonally warped). The degree of the trigonal warping of the Fermi surface differs depending on the number of layers. For monolayer graphene, the trigonal warping results from higher order terms in the $k \cdot p$ methods [40], and the effect is small in the low energy regime where transport phenomena are concerned [40] (upper left panel of Fig. 1(a)). On the other hand, bilayer graphene is expected to have significant anisotropy or a trigonally warped Fermi surface (upper right panel of Fig. 1(a)). Recently, such anisotropy in the band structure was observed by making ballistic transport measurements on graphene antidot samples [41]. Low-field magnetoresistance traces of the bilayer graphene showed commensurability magnetoresistance indicating anisotropy depending on the relative angle between crystallographic axes and the antidot lattice, while, in monolayer graphene, no such effects were observed. These results indicate that Fermi surface is significantly warped in bilayer graphene, but approximately isotropic in monolayer graphene [41].

In this work, we investigated Fermi surface anisotropy in trilayer graphene that has multiple bands by performing a ballistic transport experiment. The lower panels in Fig 1(a) show the dispersion relations and shapes of the Fermi surface in AB-stacked trilayer graphene. There are two bands, mono- and bilayer-like. The monolayer-like band, whose dispersion relation is linear, has approximately a circular Fermi surface, while the bilayer-like band, which is massive, has a larger Fermi surface area that is trigonally warped. The Fermi surface of the bilayer band is evidently more warped than that of the bilayer graphene in Fig. 1(b). We will show that ballistic transport measurements can be used to detect the monolayer-like band and the bilayer-like band that is anisotropic in plane.

Here, we briefly describe the commensurability magnetoresistance in antidot lattices. In the semi-classical picture, electrons at the Fermi level move along the circumference of the Fermi surface in a magnetic field. Therefore, the shape of the Fermi surface is closely related to that of cyclotron orbits, the electron trajectories in a magnetic field. Figure 1(b) shows schematic drawings of a Fermi surface and a cyclotron orbit at the K-point for mono- and bilayer graphene (whose crystallographic axes are as such that the $x$-axis is the zig-zag direction). Anisotropic Fermi surfaces generally result in anisotropic cyclotron orbits. Therefore, detection of an anisotropic cyclotron orbit is direct proof of an anisotropic Fermi surface. Such proof could be acquired in a ballistic transport experiment using an antidot lattice [41-42], which, as shown schematically in Fig. 1(c), is a regular array of nano-holes [42-44,46-51]. If the mean free path of the conductor is sufficiently long, the magnetoresistance shows peaks (commensurability peaks) arising from matching of the cyclotron diameter with the distance between antidots [42-44,46-50] (cases 1-3 in Fig. 1(c)). The largest peak appears when the cyclotron diameter $2R_c = 2\hbar k_f/eB$ matches the distance between the centers of the nearest antidots, *i.e.*, $2R_c = a$ (case 1). Here, $R_c$ is the cyclotron radius, $a$ is the distance between the centers, $k_f$ is the Fermi wave number, $\hbar$ is the Planck constant, $e$ is the unit charge and $B$ is the magnetic field.

The principle of detecting anisotropic cyclotron orbits with an antidot lattice sample is rather straight forward. The electron trajectories in the antidot lattice are significantly affected by the shape of the cyclotron orbits, and they determine the magneto conductivity components [41], *i.e.*, magnetoresistance traces. Therefore, observing anisotropic magnetoresistance by changing the orientation of the antidot lattice with respect to the crystallographic axes, reveals the anisotropy of the Fermi surface as shown in previous work [41]. In the case of monolayer graphene, the rotation did not change the magnetoconductivity components significantly because of the rotational symmetry. On the other hand, in the case of bilayer graphene, rotation changed them significantly [41]. Magnetoresistance without the peaks of commensurability showed clearly different behaviors depending on the angle between the crystallographic axes and the antidot lattices, while the magnetic fields of the commensurability peak remained approximately unchanged [41].

2. Sample structure and characterization

We fabricated graphene antidot lattice samples of AB-stacked trilayer graphene, whose optical micrograph is shown in Fig. 2(a). Graphene flakes were prepared by exfoliating high-quality Kish graphite with adhesive tape [1]. The number of layers was roughly determined from the color signal intensity of digitized optical images of graphene flakes [52-54]. The number of layers and stacking were further verified by examining the Landau level structure (the Shubnikov-de Haas oscillations) observed in a magnetotransport experiment. The graphene was encapsulated with hexagonal boron nitride ($h$-BN) on a $p^{++}$ −doped Si substrate covered with 300 nm of thermally oxidized $SiO_2$. The Si substrate remained conducting even at the lowest temperature and served as a back gate.

The most important point in this study is the angle of rotation of the antidot lattice with respect to the crystallographic axes of graphene. As was done in our previous work [41], a graphene flake having a clear straight edge was chosen, and the edge was used as a reference of the orientation for forming the antidot lattice (Fig. 2(b)). The straight edges were cleaved lines (CL). They were possibly disordered and were neither clear zigzag edges nor armchair edges on the microscopic scale [55]. Nevertheless, one can expect that such an edge would be a good reference of the crystallographic axes because a moiré superlattice of $h$-BN graphene heterostructures can actually be formed by using CL as references of the crystal axes [56-59]. We formed two antidot lattices on the same graphene flake; the one had a primitive vector parallel to the edge ($\theta$ =0°), and the other had one rotated by $\theta$ = 30° (Fig. 2(a)). The lattice constant (distance between the centers of the nearest antidots) was 700 nm, and the diameter of each antidot was about 200 nm.

The antidot lattices were formed by using electron beam lithography. A film of organic e-beam resist, patterned into triangular antidot lattices, was used as a mask for the reactive ion etching using a mixture of low-pressure $CF_4$ and $O_2$ gas. Electrical contacts with the graphene sample were formed by use of the edge-contact technique [60]. Resistivity was measured with the four probe method using the standard lock-in technique. Magnetic fields were applied perpendicular to the sample by using a superconducting solenoid.

Figure 2(c) shows resistivity ($\rho$) measured as a function of gate voltage ($V_g$) at liquid helium temperature ($T = 4.2$ K) and zero magnetic field ($B = 0$ T). The antidot lattice samples had an electrical mobility, $\mu = 1/|n_{tot}|e\rho$ of about $2 \times 10^4$ cm²/Vs at a high total carrier density $n_{tot}$ (Fig. 2(d)). The mobility of pristine graphene should be larger than this value because the present sample was influenced by scattering of electrons by the antidots.

3. Results

First, we verified that our sample was AB-stacked trilayer graphene by observing its Landau level structure. Figures 3 (a) and (b) show maps of longitudinal resistivity ($\rho_{xx}$) and its derivative with respect to the magnetic field ($dR_{xx}/dB$), plotted as a function of $n_{tot}$ and magnetic field ($B$). Shubnikov-de Haas (S-dH) oscillations arising from Landau levels can be seen as stripes. The structure of the Landau levels approximately resembled those in past experiments performed on the AB-stacked trilayer graphene [28-31,39]. By analyzing the filling factors at the highest magnetic field, the specific capacitance of our sample was estimated to be about $C_g = 116$ aF/μm². The conspicuous stripes that have a radial structure are principally due to the Landau levels of the bilayer-like bands. The Landau levels of the monolayer-like band appeared as beatings appearing at lower magnetic fields (indicated by the arrows in Fig. 3(a)). A characteristic feature of the AB-stacked trilayer graphene is the two different zero-mode Landau levels. The zero-mode Landau level of monolayer-like band with four-fold degeneracy can be seen (marked by m in Fig. 3(b)) at $n_{tot} \sim 0.5 \times 10^{12}$cm$^{-2}$, while the one for the bilayer-like band with eight-fold degeneracy can be seen (marked by b in Fig 3 (b)) at $n_{tot} \sim -0.5 \times 10^{12}$cm$^{-2}$.

The carrier density for each band ($n_{band}$), i.e., mono- or bilayer-like band, which was calculated from the S-dH oscillation, was also consistent with the dispersion relation of AB-stacked trilayer graphene. Figure 3(c) shows a map of the power spectra of a fast Fourier transform (FFT) of the magnetoresistance trace with respect to $1/B$. The frequency of the FFT is scaled so that the peaks of the FFT directly relate to $n_{band}$ for each band resulting in S-dH oscillations. The measured carrier density for each band could be approximately reproduced by a band calculation based on the tight binding model utilizing the SWMcC parameters of graphite, although the carrier density for the monolayer was slightly larger than in the experiment. These features proved that the sample was AB-stacked trilayer graphene.

Peaks due to commensurability magnetoresistance are visible at for $B < 1$ T in Fig. 3(b) (indicated by arrows). These peaks are absent from the measurements of the samples without the antidot lattice [28-30, 39]. The commensurability peaks can be more clearly seen in the magnetoresistance traces. Figures 4(a) and (b) show the results for the antidot lattice with $\theta = 0°$ and $30°$, respectively. As shown from bottom to top, the gate voltage was varied from -50 to 50 V in 10 V steps. Commensurability peaks (shown by arrows) arising from the antidot lattice are clearly visible in the traces. In addition, at magnetic fields higher than $B > 0.5$ T, small oscillations with fast and slow periods appeared. These can be identified to be S-dH oscillations for the bilayer-like and monolayer-like bands, by comparing them with those of the AB-stacked trilayer graphene sample without the antidot lattice.

The main commensurability peak appears at a magnetic field where the cyclotron diameter matches the distance between the center of the nearest antidots, $i.e.$, $2R_c = a$. Using a simple formula, $k_f \approx \sqrt{\pi n_{band}}$, the magnetic field due to the main commensurability peak can is given by $B_p = 2\hbar\sqrt{\pi n_{band}}/ea$, and this yields $B_p \approx 0.51$ T for $V_g = -50$ V for the bilayer-like band, which agrees fairly well with the largest commensurability peaks (b1) for $V_g = -50$ V in Figs. 4(a) and (b). Higher order peaks possibly arising from the next nearest antidots (labeled b2, case 2 in Fig. 1 (c)) are also visible in the magnetoresistance trace.

The monolayer-like band, on the other hand, has a carrier density of about $0.3 \times 10^{12} cm^{-2}$ for $V_g = -50$ V, and the commensurability peak is expected to appear around $B = 0.16$ T. The magnetic field is close to that of the peak appearing after a sharp increase from zero magnetic field (m1). These peaks have not been observed in experiments on mono- and bilayer graphene antidot samples [41,49-50], $i.e.$, they are characteristic of AB-stacked trilayer graphene. In monolayer graphene, the low-field magnetoresistance showed monotonic behavior, and slightly negative magnetoresistance was observed [41,49-50]. In bilayer graphene, magnetoresistance without peaks due to commensurability magnetoresistance showed monotonic behavior; it increased or decreased depending on the orientation of the antidot lattice with respect to the crystallographic axes [41].

The values of $B_p$ were further analyzed. Figure 5 (a) summarizes the measured $B_p$ for the main commensurability peaks. Figure 5 (b) shows the estimated $n_{band}$ for mono- and bilayer-like bands (Fig. 5(b)). The measured values of $B_p$ can approximately be explained by the estimations made from the carrier density.

Next, let us focus on the in-plane anisotropy. The magnetoresistance traces for $\theta = 30°$, shown in Fig. 4 (b), have similar peak structures arising from commensurability, but also have two features different from the case of $\theta = 0°$. One is that the peaks for the main commensurability (indicated by down arrows) are considerably smaller than the case of $\theta = 0°$. The other is that higher order commensurability peaks b2 are less visible in the data with $\theta = 30°$. These features are qualitatively similar to those of bilayer graphene so they should originate from the Fermi surface anisotropy of the bilayer-like band [41].

To investigate the anisotropic behavior of the commensurability magnetoresistance, we performed numerical simulations based on a semi-classical model. The conductivity components of each band (mono- or bilayer-like band) were calculated by [61],

$$\sigma_{ij} = A \int_0^\infty <v_i(0)v_j(t)> \exp(-t/\tau)\ dt, \qquad (3)$$

where $v_i$ and $v_j$ are group velocities in the $i$ and $j$ directions, $\tau$ is the relaxation time, and $A$ is a constant. $<\cdots>$ indicates the average over possible initial states. We numerically calculated all the conductivity components, i.e., $\sigma_{xx}$, $\sigma_{xy}$, $\sigma_{yx}$, and $\sigma_{yy}$, for each band and added them together with a suitable weight to obtain the conductivity components of the whole system. Then, the resistivity components were calculated by tensor inversion. To evaluate eq. (3), we used a model Fermi surface whose analytic formula is given in polar coordinates $(k_f, \theta)$ by

$$k_f = k_0\big(1 + \alpha\ \cos(3\theta) + \beta\ \cos(6\theta)\big). \qquad (4)$$

Here, $\alpha$ and $\beta$ are parameters describing trigonal warping of a Fermi surface, and the $x$-axis is the direction of the zigzag edge. These parameters were estimated by fitting eq. (4) to numerically calculated energy contour of the dispersion relations. For $n \approx 3 \times 10^{12}$ cm$^{-2}$, they were estimated to be $\alpha = 0.15 \sim 0.2$ and $\beta \approx 0.06$ for the bilayer-like band, and $\alpha \approx 0$ and $\beta \approx 0$ for the monolayer band.

Figures 6 (a) and 6 (b) show the numerically calculated resistivities for $\alpha = 1.5$. Here, $R_{c2L}$ is the cyclotron radius of the bilayer-like band, and therefore, $a/R_{c2L}$ is proportional to the magnetic field. The ratio of the carrier density of the monolayer-like band ($n_{1L}$) and that of the bilayer-like band ($n_{2L}$) was taken to be $n_{1L}/n_{2L} = 0.1$, which approximates the experimental values. The mean free path $l_f$ was chosen to be $1.5a$, and the weights of the conductivity component ($\sigma_{1L}/\sigma_{2L}$) is served as adjustable parameters. The experimentally measured magnetoresistances were qualitatively reproduced when $\sigma_{1L}/\sigma_{2L} = 0.5 \sim 1$. The maximum resistance peak around $a/R_{c2L} \approx 2$ (b1) is the main peak of the commensurability for the bilayer-like band. A peak appearing around $a/R_c \approx 1.2$ (b2) are for commensurability associated with the next nearest neighbor antidot. The peak appearing at $a/R_c \approx 0.5$ (m1) is the main commensurability peak for the monolayer-like band. The simulation also qualitatively reproduced the anisotropy with respect to $\theta$; magnetoresistance (without the peaks for the commensurability) for $\theta = 0°$ is larger than that for $\theta = 30°$, as in the experiment, which would show that $\theta = 0°$ is the zigzag direction in our sample. This anisotropic behavior should arise from the anisotropic Fermi surface of the bilayer-like band, which is similar to the case of the bilayer graphene [41]. In addition, anisotropy can be seen in other points. The commensurability peak (b2) is rather weak for $\theta = 30°$, but clearly visible for $\theta = 0°$, as in the experiment (Figs. 4 (a) and 4 (b)). Moreover, the magnetic field of the main commensurability peak for $\theta = 0$ is slightly larger than that for $\theta = 30°$.

Next we turn our attention to the steep increase in resistance from zero magnetic field to peak m1 (at $a/R_{c2L} \approx 0.5$), which qualitatively reproduced the positive magnetoresistance in the experiment. Note that such positive and peaked magnetoresistance was not observed in the experiment (and numerical simulations) in antidot lattice samples of mono- and bilayer graphene [41, 49-50]. Therefore, it should also originate from the multiband transport properties. It is known that if multiple bands contribute to electronic conduction, unbalanced longitudinal and Hall conductivities result in a positive magnetoresistance at low magnetic fields, whereas no semi-classical magnetoresistance appears for electron systems with a single isotropic band.

Although the carrier density of the monolayer band is about an order smaller than that of the bilayer-like band, the monolayer band was found to contribute to the total conductivity significantly because the simulation indicated $(\sigma_{1L}/\sigma_{2L})$ is of order 1. This would be because the monolayer band is a massless band and has a large Fermi velocity $(v_f)$. $\sigma_{1L}$ and $\sigma_{2L}$ can be roughly estimated using the relation, $\sigma \sim Ce^2 N(0) v_f l_f$, where $N(0)$ is the density of states at the Fermi level, and $C$ is a constant of order 1. $N(0)$ and $v_f$ could be calculated from the dispersion relation by averaging $N(E) = dn/dE$ and $|\vec{v}| = |(1/\hbar)(\nabla_k E)|$ over angles. For $n_{tot} \approx 3.0 \times 10^{12}$ cm$^{-2}$, $(\sigma_{1L}/\sigma_{2L})$ was estimated to be $0.35 - 0.45$, which is the same order of magnitude as the estimation from the magnetoresistance curves. This explains why the monolayer-like band still significantly contributes to the total conductivity even though the carrier density of the monolayer-like band is about an order smaller than that of the bilayer-like band.

## Discussion

Although ballistic transport has been studied in various systems, most of the studies used systems consisting only of a single band with a simple electronic structure. The choice of a simple structure was because ballistic transport experiments require samples with sufficiently long mean free paths, and therefore, suitable materials are naturally limited to high-quality two-dimensional semiconductors grown by molecular beam epitaxy, etc. As far as the authors know, ballistic transport has only been studied in a multiband system by conducting a magnetofocusing experiment on AB-stacked trilayer graphene [62]. Here, magnetofocusing is similar to commensurability magnetoresistance in antidot lattices in that they both originate from matching of the cyclotron diameter with geometrical structures. However, antidot lattice experiments have the advantage in being able to observe the anisotropic band structure because an electron's collision with the antidot is strongly dependent on the symmetry of the cyclotron orbit that of the antidot lattice [41].

The effect of trigonal warping of the band structure of graphene has been discussed from various points of view. Trigonal warping has a significant effect on the dephasing rate of the weak localization [63-64]. It also affects the Landau level structures in multilayer graphene. The crossings of the Landau levels were found to be significantly different from the case without trigonal warping [27]. Furthermore, trigonal warping was found to significantly influence the low-energy band structure, in particular in the presence of a perpendicular electric field. Here, mini-Dirac cones formed at the bottoms of the bilayer-like band of multilayer graphene when it was placed in a perpendicular electric field [65]. Mini-Dirac cones have been observed in experiments on intrinsic resistance peaks (ridges) that appear in the dependence of the bottom gate voltage dependence on resistance [16,34,37-38]. However in-plane anisotropy of the band structure can only be successfully detected with ballistic transport by using antidot lattices.

So far various methods have been used to study the electronic band structure of materials. Optical methods, *eg*, photoelectron spectroscopy and infrared spectroscopy, are suitable for studying the band structure in a relatively wide range of energies. As for the low-energy band structure where transport phenomena appear, the Shubnikov-de Haas effect, angular dependent magnetoresistance oscillations [66-69] and intrinsic resistance peaks, etc. can be used. Ballistic electron transport experiments might be another way to probe the low-energy band structure in two-dimensional materials.

Conclusion

We observed multiband ballistic transport in an antidot lattice of AB-stacked trilayer graphene. It was found that the low-field magnetoresistance had a characteristic structure that qualitatively differed from the magnetoresistance of mono- and bilayer graphene antidot-lattice samples. Besides clear commensurability peaks arising from the bilayer-like band, we observed peaks possibly due to commensurability for the monolayer-like band at a low magnetic field. Magnetoresistance without the peaks due to the commensurability showed a clear dependence on the orientation of the antidot lattice, which would indicate the anisotropic features of the bilayer-like band. Moreover, although the carrier density of the monolayer-like band was an order smaller than that of the bilayer-like band, a monolayer-like band significantly contributed to the conductivity. The experimental magnetoresistance was approximately reproduced by the numerical simulations based on semi-classical transport. These findings suggest that ballistic experiments might be a new method of studying the electronic band structure of two-dimensional materials.

## Acknowledgements

This work was supported by a grant KAKENHI No.25107003 from MEXT Japan.

**Appendix A    Model Fermi surface and numerical simulation of magnetoresistance**

To evaluate equation (3), one needs to calculate the velocity, and this requires information about the dispersion relation of the electron system. We considered two cases, a linear, and a parabolic dispersion relation. If the energy $E$ is given by $E = \hbar \vec{v} \cdot \vec{k}$, where $\vec{v}$ is the velocity, the angular dependence of $|\vec{v}|$ is

$$|\vec{v}| \propto 1/(1 + \alpha \cos(3\theta) + \beta \cos(6\theta)). \tag{5}$$

In the case of a parabolic dispersion, the energy is given by a simple formula, $E = \hbar k(\theta)^2/2m(\theta)$, where $m(\theta)$ is the band mass which is dependent on the crystallographic orientations. Accordingly, the angular dependence of $m(\theta)$ is proportional to

$$m(\theta) \propto 1/\bigl(1 + \alpha \cos(3\theta) + \beta \cos(6\theta)\bigr)^2. \tag{6}$$

For this case also, the group velocity satisfies eq. (5). The electron orbitals in $k$-space can be calculated by solving the semi-classical equation,

$$\hbar \frac{d\vec{k}}{dt} = -e\vec{v} \times \vec{B}, \tag{7}$$

where $\vec{B}$ is magnetic field. The orbit in the real space can be obtained by integrating the group velocity. The cyclotron orbit is interrupted by the collision with the antidot. In the simulation, we assumed specular scattering. In an actual graphene antidot lattice to be used in experiments, the circumference of the antidots should be substantially disordered on the atomic scale so that the scattering will be diffusive. However, the nature of the scattering does not change the important features of the calculated commensurability magnetoresistance [41,70-71].

Figure captions

Fig. 1 Cyclotron orbit and antidot lattice.
 (a) Numerically calculated energy contour of the dispersion relation for mono-, bi- and AB-stacked trilayer graphene (for a total carrier density $n_{tot} \approx 3.0 \times 10^{12}$ cm$^{-2}$ ). The dispersion relation for AB-stacked trilayer graphene is shown at the bottom left. The dashed line indicates $E = 67$meV for which the energy contour for the trilayer graphene was calculated. The calculations are based on the effective mass approximation. Slonczewski-Weiss-McClure (SWMcC) parameters of graphite ($\gamma_0 = 3.16$ eV, $\gamma_1 = 0.39$ eV, $\gamma_2 = -0.02$ eV, $\gamma_3 = 0.3$ eV, $\gamma_4 = 0.044$eV, $\gamma_5 = 0.038$ eV, and $\Delta_p = 0.037$ eV) were used. (b) Schematic drawings of the Fermi surface and cyclotron orbit. Electrons in a circular Fermi surface (1L) move along a circular orbit in magnetic fields while trigonally warped Fermi surface (2L) results in a trigonally warped cyclotron orbital. (c) Schematic illustration of origin of the commensurability magnetoresistance. Cyclotron orbits in triangular antidot lattice are shown for different commensurability conditions (indicated by 1-3). A magnetoresistance peak appears at magnetic fields under which the cyclotron diameter matches the distance between the centers of the antidot.

Fig. 2  Fabrication and characterization of trilayer graphene antidot sample.
(a) Optical micrograph of a graphene antidot sample before connecting electrical leads (left). Numbers indicate probe numbers. Graphene was h-BN encapsulated. CL indicates the direction of the cleaved line of the graphene flake. Antidot lattices with two different orientations were formed on the graphene flake.  One had a primitive vector parallel to the cleaved line; the other was that rotated by 30°. (b) Optical micrograph of an *h*-BN encapsulated graphene flake before patterning into a Hall bar. G is graphene, hBN is *h*-BN. (c)  $V_g$-dependence of resistivity $\rho$ of the graphene device. (d) $V_g$ dependence of electrical mobility $\mu (= 1/n_{tot} e\rho)$, where $n_{tot}$ is the total carrier density.

Fig. 3 Landau fan diagrams for antidot lattice samples of AB-stacked trilayer graphene.
(a) Map of $\rho_{xx}$ in antidot sample with $\theta = 0$ as a function of $n_{tot}$ and magnetic field ($B$). Arrows indicate the S-dH effect arising from the monolayer-like band. $T = 4.2$ K. The Landau level structure for the sample with $\theta = 30°$ was approximately the same (not shown). (b) Map of $d\rho_{xx}/dB$. Arrows inidicate commensurability peaks arising from antidot. The labels m and b indicate zero-mode Landau levels of the mono- and bilayer-like bands. (c) Map of Fourier spectra for the $1/B$ dependence of magnetoresistance. Here, the frequency of the FFT was converted into the carrier density ($n_{osc}$) of the electronic band that causes the S-dH oscillation component. Theoretical values, numerically calculated with the tight-binding model using SWMcC parameters of graphite, are indicated by dashed lines. Mono- and bilayer-like bands are labeled m and b.

Fig. 4

Traces of normalized longitudinal magnetoresistance ($\rho_{xx}(B)/\rho(0)$) of the antidot samples plotted for different gate voltages. From bottom to top, $V_g$ was varied from $-50$ to 50 V in 10 V steps. Panel (a) is for $\theta = 0$, where probes 2 and 3 were used as voltage probes, and probes 1 and 6 were used as current probes. Panel (b) is for $\theta = 30°$, where probes 4 and 5 were used as voltage probes. Data are offset by 0.075. Magnetoresistance is normalized by the resistance at zero magnetic field. $T = 4.2$ K. Downward arrows indicate main commensurability peaks (labeled b1) corresponding to case 1 in Fig. 1(c) for the bilayer-like band. Peaks (only shown for the bottom data with b2) indicate the peak arising from case 2 in Fig. 1(c). Peaks (only shown for the bottom data with m1) indicate the main commensurability peak of the monolayer-like band.

Fig. 5 Analysis of $B_p$

(a) $V_g$ dependence of $B_p$ for different peaks. Solid triangles and solid squares are, respectively, $B_p$ for the main commensurability peak for bilayer- and monolayer-like bands in the magnetoresistance trace. Open triangles and open squares are $B_p$ calculated from carrier densities determined from the S-dH effect. The lines are guides for the eye. (b) Carrier densities $n_{band}$ in mono- and bilayer-like bands, determined from the S-dH effect as a function of total carrier density $n_{tot}$ (solid triangles and solid squares, respectively). Lines are carrier densities calculated from the band calculation shown in Fig 1(a). Labels m and b denote the mono- and the bilayer-like band.

Fig. 6 Simulation of resistivity for trilayer graphene antidot lattice

Numerically calculated traces of magnetoresistivity for different values of $\sigma_{1L}/\sigma_{2L}$. Panels (a) and (b) are for $\theta = 0°$ and 30°. $\alpha = 0.15$ and $\beta = 0.06$ for the bilayer-like band. $\alpha = 0$ and $\beta = 0$ for the monolayer-like band. Mean free path $l_f$ is was taken to be $l_f/a = 1.5$ for both bands. $d/a = 0.2$. b1 and m1 show main commensurability peaks of the bilayer-like and monolayer-like band. b2 show the commensurability peak for the next-nearest neighbor antidots.

(a)

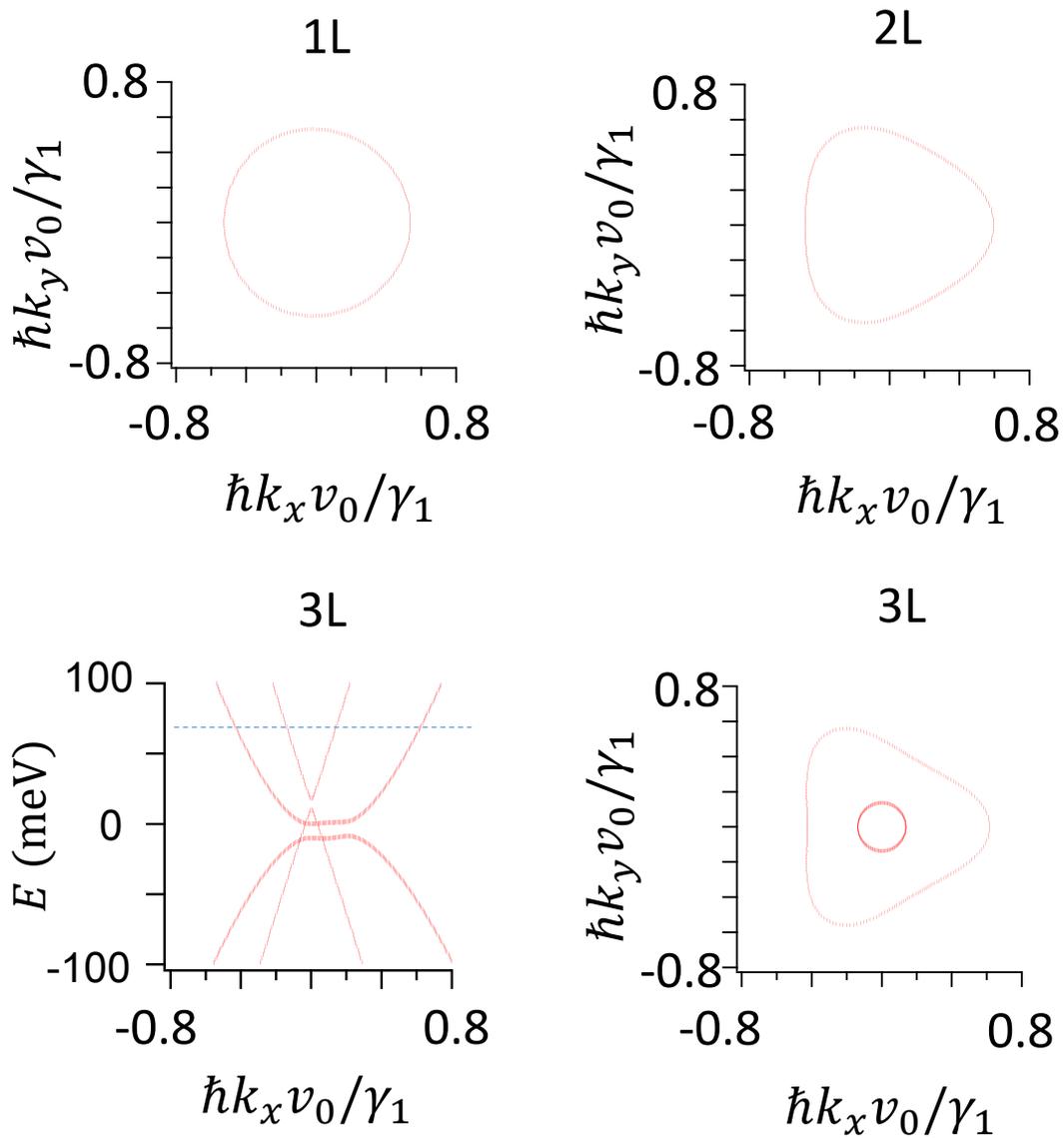

(b)

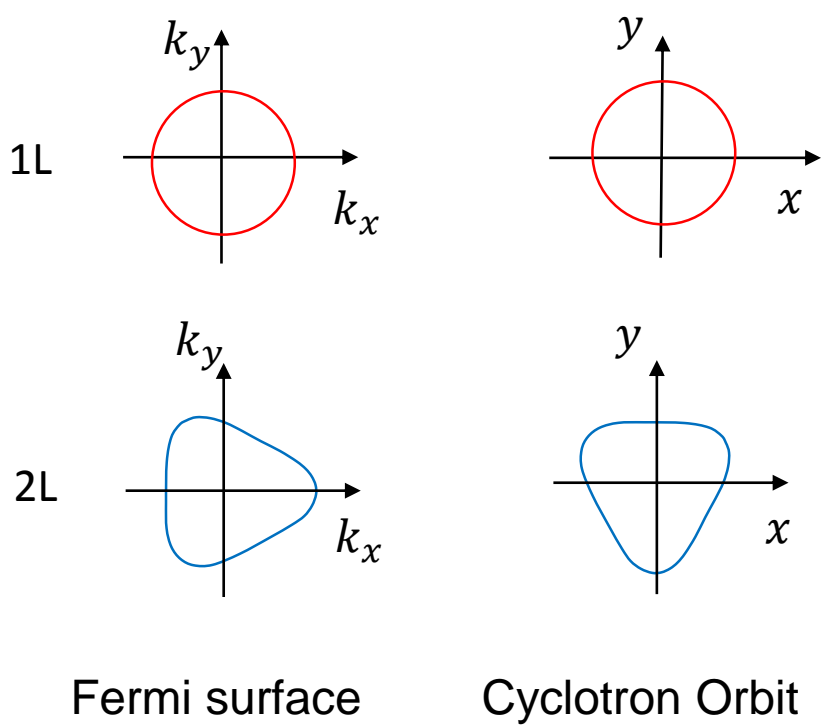

Fermi surface      Cyclotron Orbit

(c)

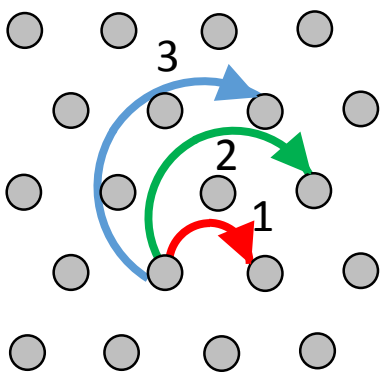

Fig. 2

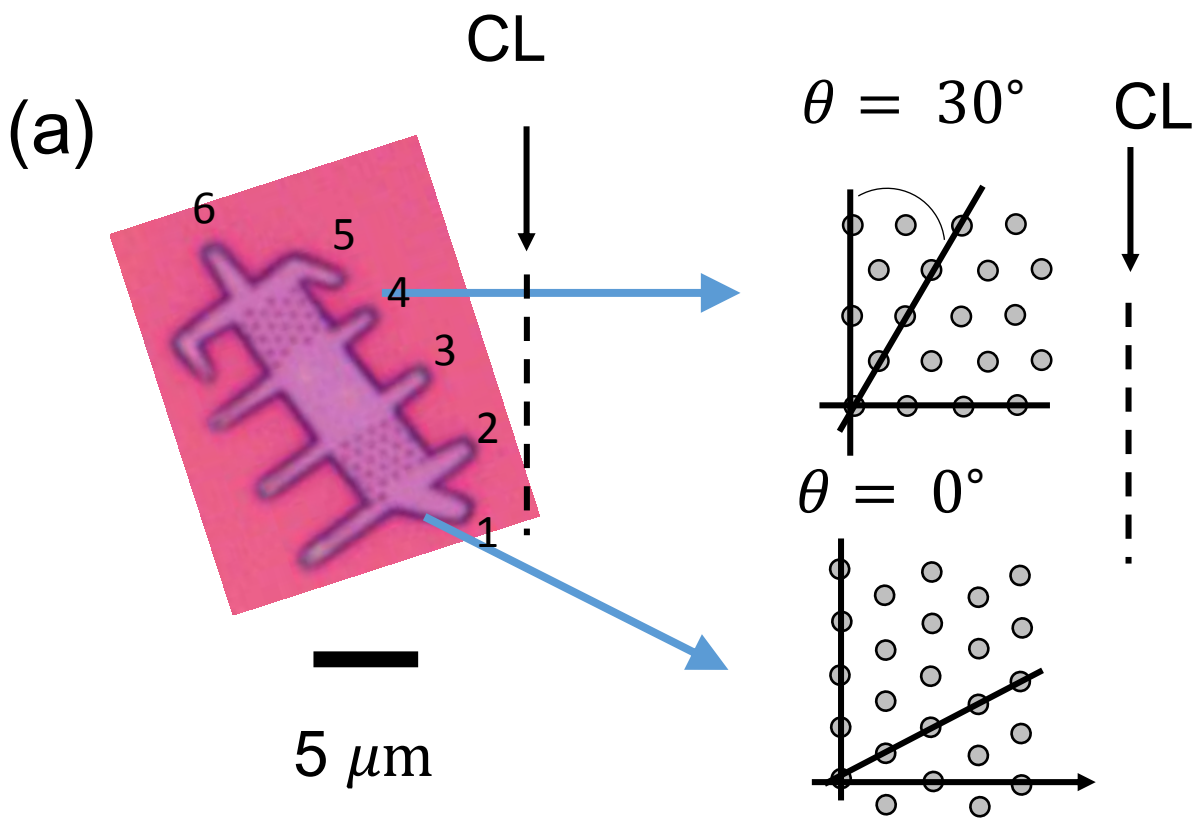

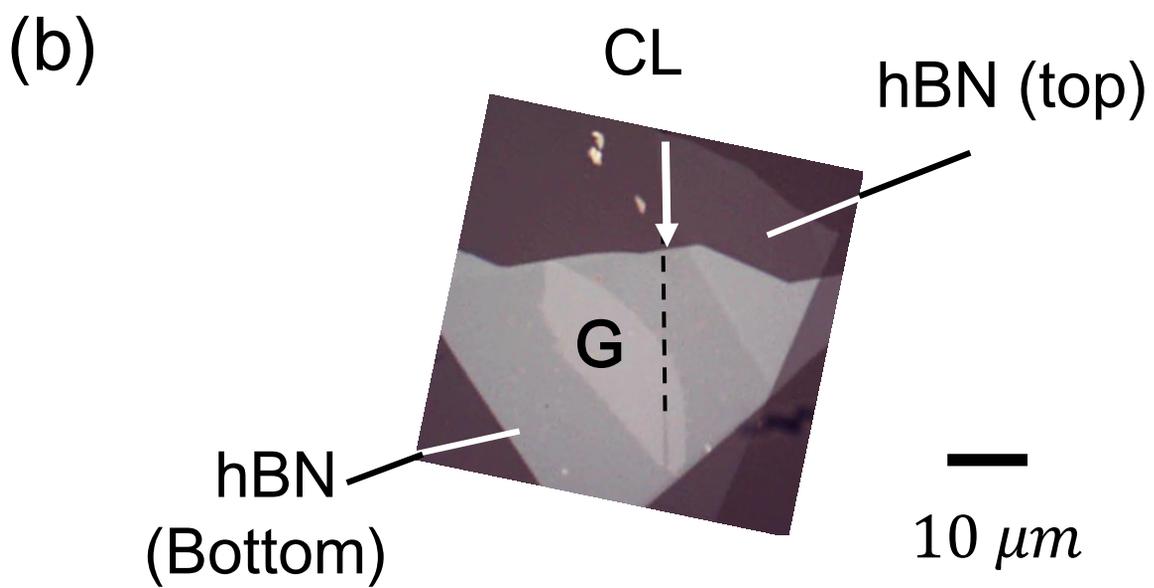

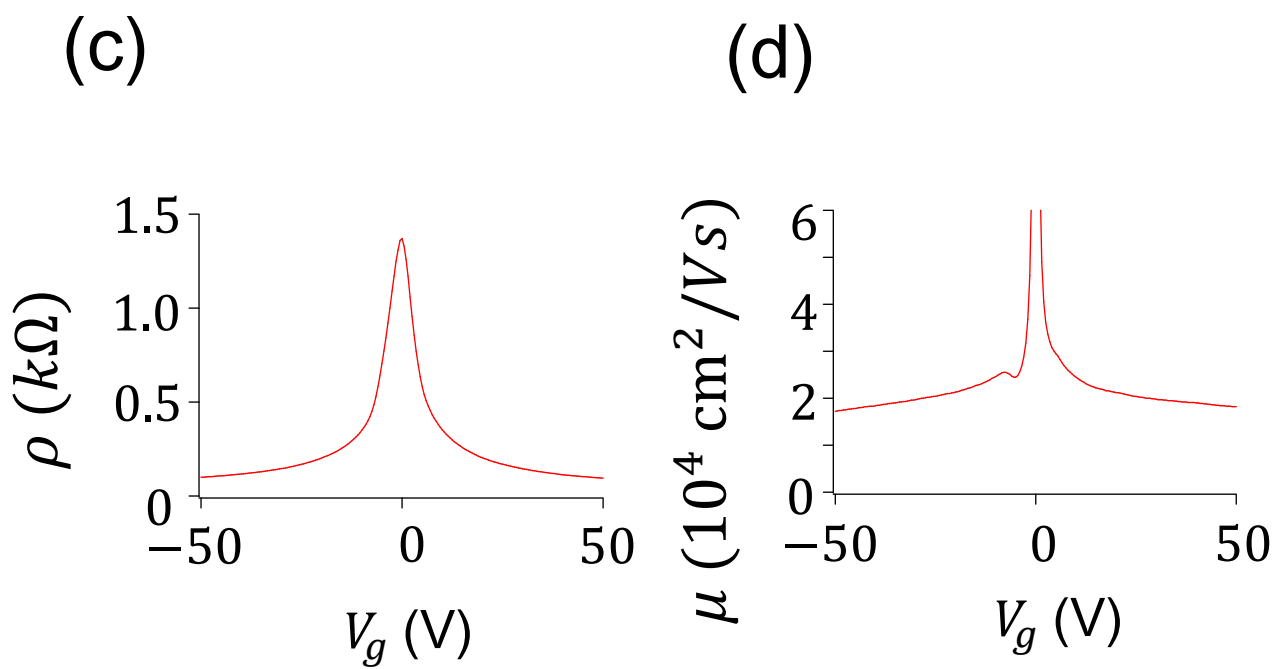

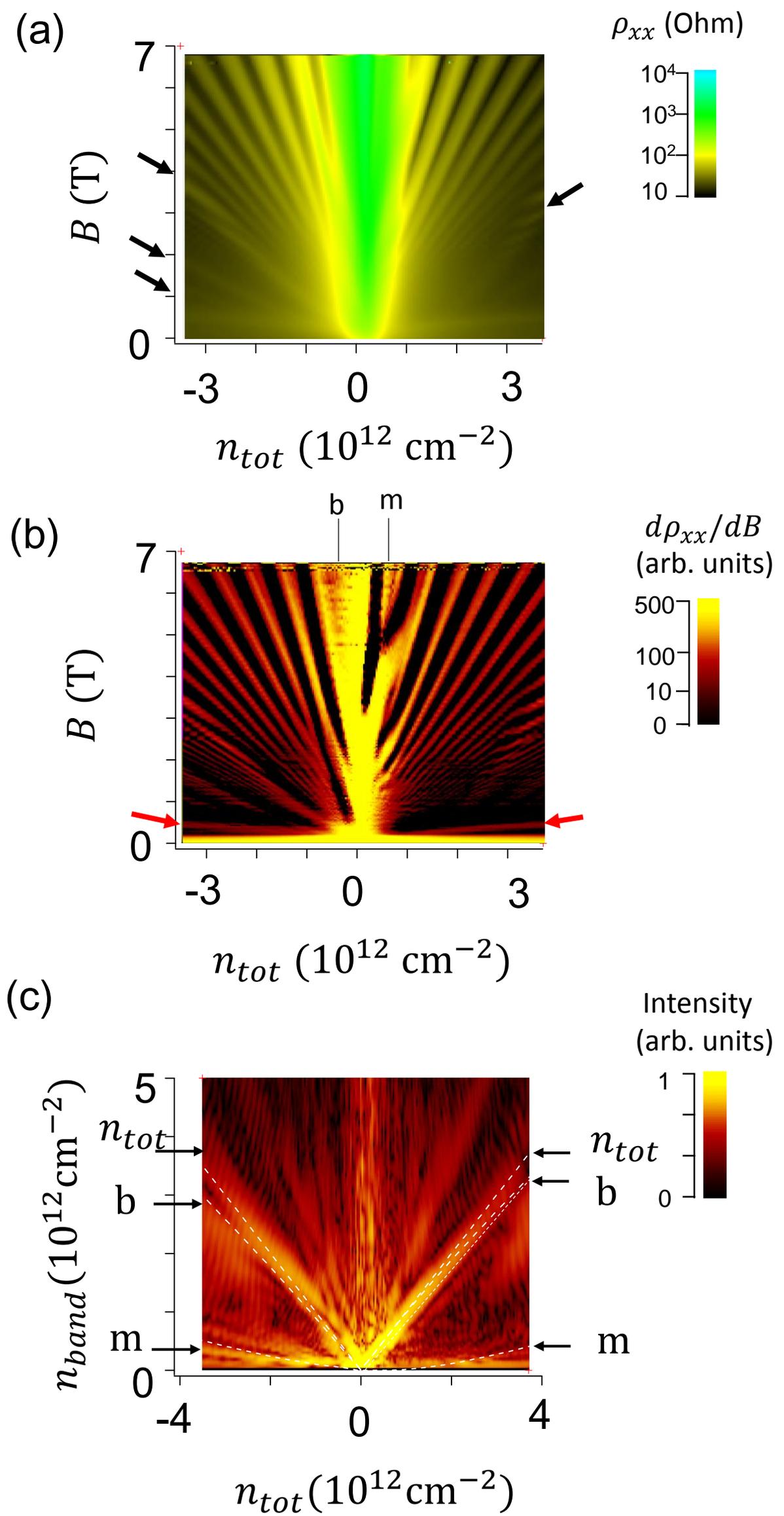

Fig. 3

Fig. 4

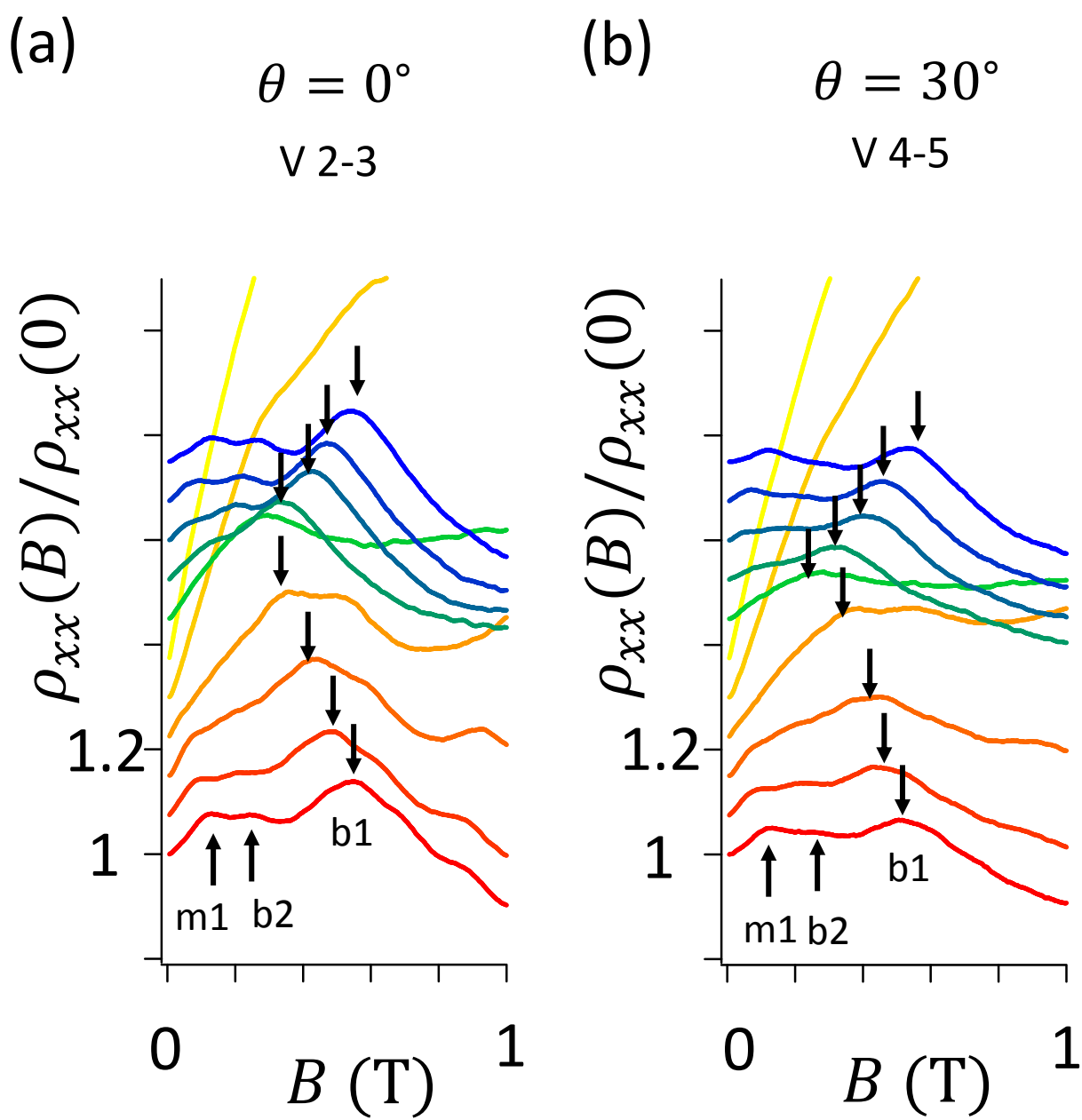

Fig. 5

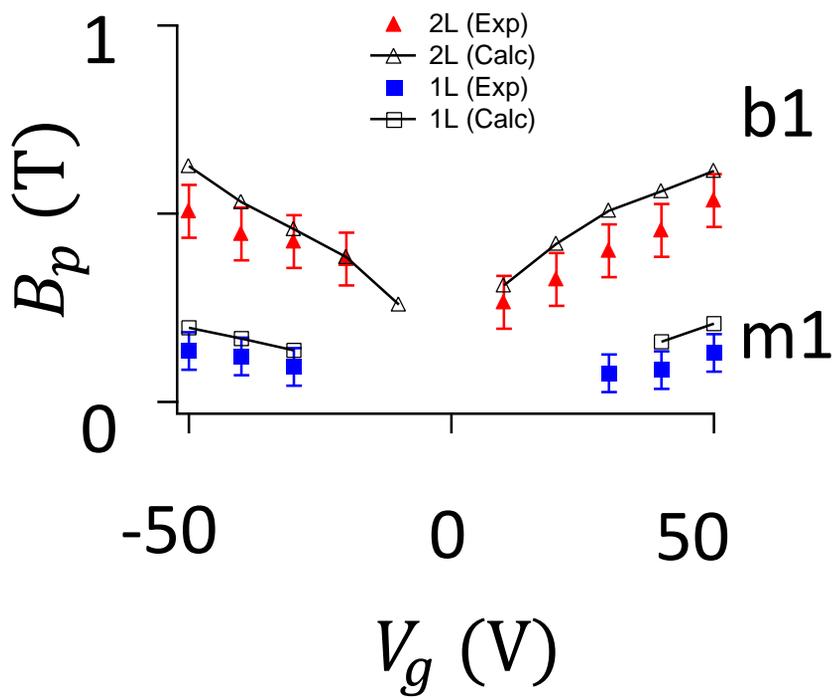

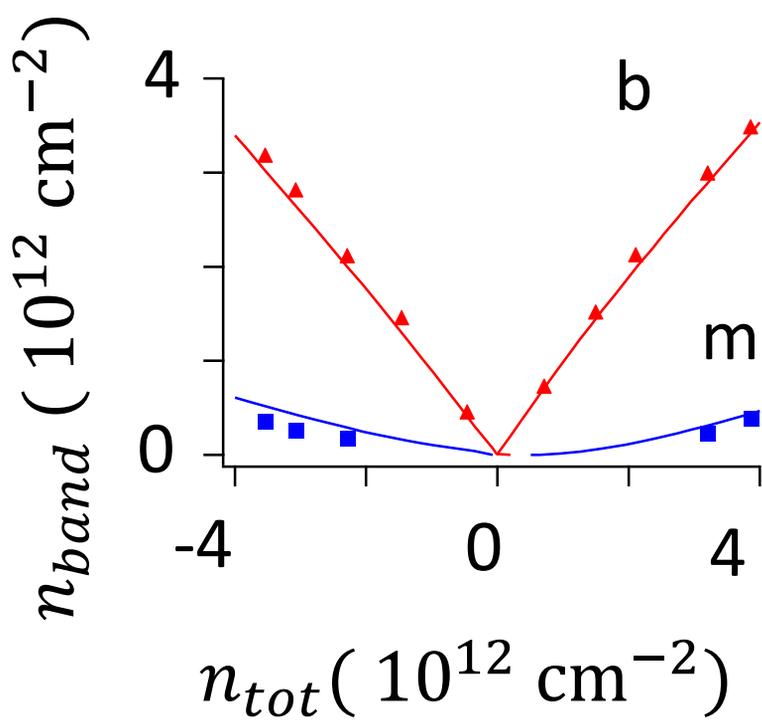

Fig. 6

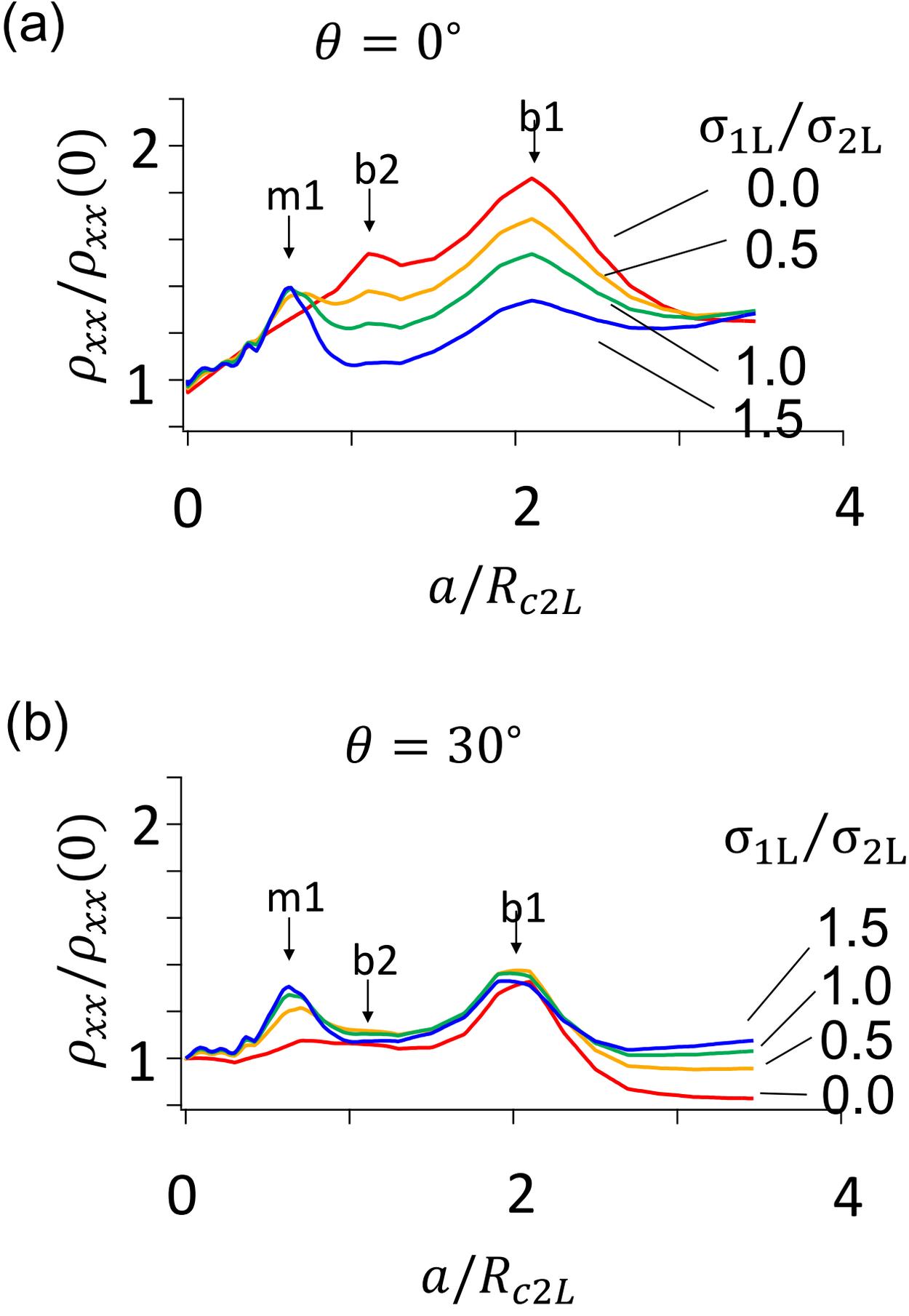